\documentclass[aps,twocolumn,floats,pre]{revtex4-1}
\usepackage{graphics,graphicx,epsfig}
\usepackage{amssymb,color}
\usepackage{amsmath}
\usepackage{epsf,epstopdf,wrapfig}

\begin{document}
\title{Spike-timing prediction with active dendrites}
\author{Richard Naud$^1$, Brice Bathellier$^2$ and Wulfram Gerstner$^3$}
\affiliation{$^1$ Department of Physics, University of Ottawa, 150 Louis Pasteur, ON, K1N 6N5, Canada. \\
$^2$ Unit of Neuroscience Information and Complexity (UNIC) CNRS UPR-3239, 1 av. de la Terasse, Gif-sur-Yvette, 91198, France.\\
$^3$ School of Computer and Communication Sciences and School of Life Sciences, Ecole Polytechnique Federale de Lausanne, Building AAB Lausane-EPFL, 1015, Switzerland.\\
}
\date{\today}

\begin{abstract}
A complete single-neuron model must correctly reproduce the firing of spikes and bursts.   We present a study of a simplified model of deep pyramidal cells of the cortex with active dendrites.  We hypothesized that we can model the soma and its apical tuft with only two compartments, without significant loss in the accuracy of spike-timing predictions. The model is based on experimentally measurable impulse-response functions, which transfer the effect of current injected in one compartment to current reaching the other.  Each compartment was modeled with a pair of non-linear differential equations and a small number of parameters that approximate the Hodgkin-and-Huxley equations.  The predictive power of this model was tested on electrophysiological experiments where noisy current was injected in both the soma and the apical dendrite simultaneously. We conclude that a simple two-compartment model can predict spike times of pyramidal cells stimulated in the soma and dendrites simultaneously. Our results support that regenerating activity in the dendritic tuft is required to properly account for the dynamics of layer 5 pyramidal cells under in-vivo-like conditions.
  \end{abstract}
\maketitle

\section{Introduction}
\label{sec-intro}
Partially neglected for a long time, dendrites have been recently shown to treat synaptic input in a surprising variety of modes\cite{Stuart2007b}. One particularly striking example is found in pyramidal cells of deep cortical layers. In these cells, a coincidence between a back-propagating action potential and dendritic input can trigger voltage-sensitive ion channels situated on the apical dendrite more than 300~$\mu$m from the soma \cite{Larkum1999a,Larkum2001a}. The somatic membrane potential increases only after  the activation of dendritic ion channels. This often resulting in a burst of action potentials.  Bursts in these cells can therefore signal a coincidence of input from the soma (down) with inputs in the apical dendrites (top). Such top-down coincidence detection is one computation that is attributed to dendritic processes. Other allegedly dendritic computations include subtraction \cite{Gabbiani2002a}, direction selectivity \cite{Taylor2000a}, temporal sequence discrimination \cite{Branco2010a}, binocular disparity \cite{Archie2000a}, gain modulation \cite{Larkum2004a} and self-organization of neuron networks \cite{Legenstein2011a}.

Models of large pyramidal neurons that are active at the tuft of their apical dendrites were first described by Traub \textit{et al.} (1991) \cite{Traub1991a} for the hippocampus. This model of the large CA3 pyramidal neurons included voltage-dependent conductances on the dendrites.  It is a model based on the Hodgkin-Huxley description of ion channels. Cable properties of dendrites are taken into account by segmenting the dendrite into smaller compartments. The resulting set of equations is solved numerically. A simplified version of this model was advanced by Pinsky and Rinzel (1994) \cite{Pinsky1994a}. They have reduced the model to a dendritic compartment and a somatic compartment connected by an effective conductance.  The model has a restricted set of five ion channels and accounts for bursting of CA3 pyramidal cells. 

Models specific to deep cortical cells have been described by extending the approach of Traub \textit{et al.} (1991). Schaefer \textit{et al.} (2003) \cite{Schaefer2003a} used morphological reconstruction to define compartments. This model could reproduce the top-down coincidence detection.  

Using a simplified approach similar to Pinsky and Rinzel (1994) \cite{Pinsky1994a}, Larkum \textit{et al.} (2004) \cite{Larkum2004a} have modelled dendrite-based gain modulation. The parameters in the model could be tuned to quantitatively reproduce the firing rate response of layer 5 pyramidal cells stimulated at the soma and the dendrites simultaneously. Larkum \textit{et al.} (2004) concluded that a two-compartment model was sufficient to explain the time-averaged firing rate.

A more stringent requirement for neuron model validation, however, is to predict spike times \cite{Keat2001a,Pillow2005a,Jolivet2006a, Jolivet2008a, Jolivet2008b, Gerstner2009a}. Given the low spike-time reliability of pyramidal neurons, spike time prediction is compared to the intrinsic reliability \cite{Jolivet2006a}. This approach can be seen as predicting the instantaneous firing rate \cite{Naud2011a}. Generalized integrate-and-fire models can predict instantaneous firing rate of layer 5 pyramidal neurons with substantial precision\cite{Jolivet2008a,Naud2009a,Gerstner2009a} in the absence of dendritic stimulation.  The question remains whether a neuron model can predict the spike times of layer 5 pyramidal neurons when both the dendrites and the soma are stimulated simultaneously. 

We present a study of a simplified model of layer 5 pyramidal cells of the cortex with active dendrites.  Following Larkum \textit{et al.} (2004) \cite{Larkum2004a}, we hypothesized that we can model the soma and its apical tuft with two compartments, without significant loss in the accuracy of spike-timing predictions. We introduce experimentally measurable impulse-response functions \citep{Segev1995a}, which transfer the effect of current injected in one compartment to current reaching the other. The impulse-response functions replace the instantaneous connection used in previous two-compartment models \cite{Pinsky1994a,Larkum2004a} and acts as a third, passive, compartment.  Each compartment was modeled with a pair of non-linear differential equations with a small number of parameters that approximate the Hodgkin-and-Huxley equations.  The predictive power of this model was tested on electrophysiological experiments where noisy current was injected in both the soma and the apical dendrite simultaneously \citep{Larkum2004a}.

\section{Methods}
\label{sec-met}

Methods are separated in four parts. First we present the model, second the experimental protocol, then fitting methods and finally the analysis methods. 

\begin{figure}[h!]
\centering
\includegraphics[width = 0.23\textwidth, height = 3in]{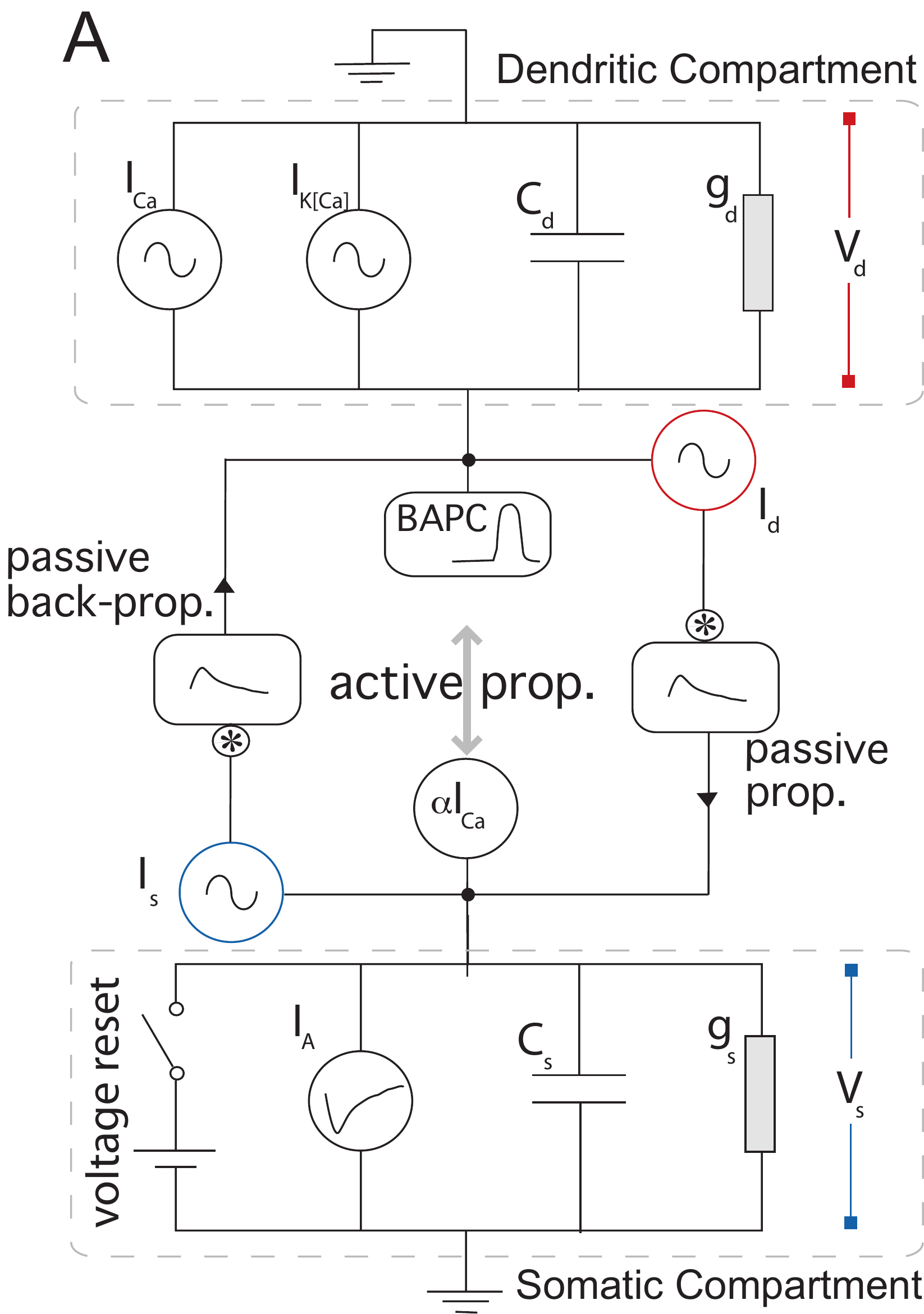}
 \includegraphics[width = 0.23\textwidth, height = 3in]{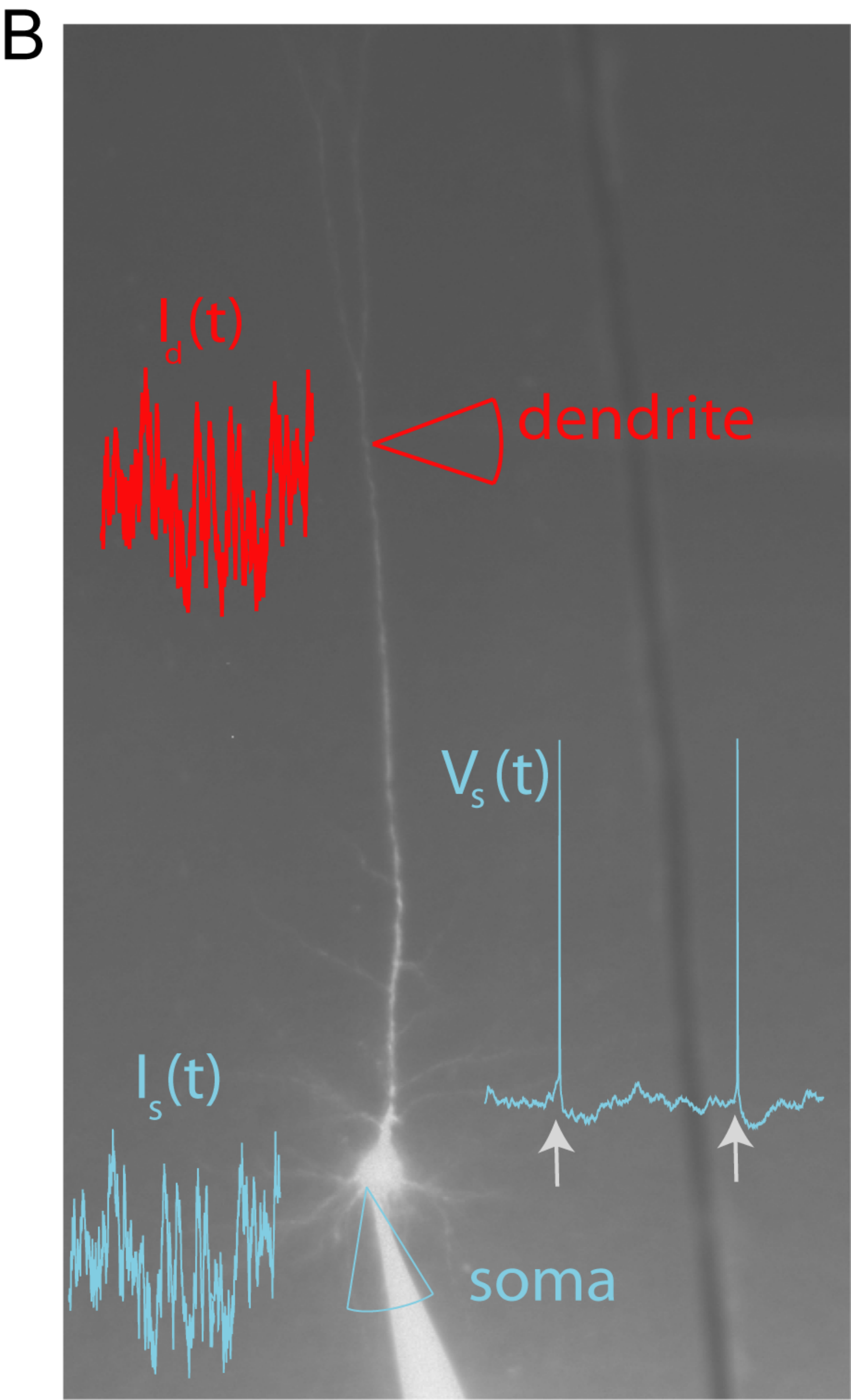}
 \caption{ Schematic representation of the two-compartment  model.  {\bf A} Somatic and dendritic compartment communicate through passive and active propagation.  Passive communication filters through a convolution (denoted by an asterisk) the current injected in the other compartment. Active communication in the soma introduces a perturbation proportional to the dendritic current $I_{Ca}$. Active communication to the dendrites introduces a stereotypical back-propagating action potential current (BAPC). The somatic compartment has spike-triggered adaptation and a moving threshold. The dendritic compartment has an activation current and recovery current. {\bf B} Associated experimental protocol with current injection both in soma and apical dendrite of layer 5 pyramidal cells of the rat somato-sensory cortex. Variables are defined in the main text.}\label{ModelSchematic}
\end{figure}
\subsection{Description of the Model}\label{ModDescrp}
Fig. \ref{ModelSchematic} shows a schematic representation of the two-compartment model.  In details, the model follows the system of differential equations:
\begin{eqnarray}
C_s\frac{dV_s}{dt} & = & -g_s(V_s -E_s) +  \alpha m + I_s \nonumber \\ 
&&  + \sum_{\{\hat{t}_i\}}I_A(t-\hat{t}_i) + \epsilon_{ds} * I_d \\
C_d \frac{dV_d}{dt} & = & -g_d(V_d - E_d)  + g_1m + g_2x + I_d \nonumber \\
&&+ \sum_{\{\hat{t}_i\}}I_{BAP}(t-\hat{t}_i) + \epsilon_{sd}*I_s \\
\tau_m \frac{dm}{dt} & = & \frac{1}{1+ \exp\left(-\frac{V_d-E_m}{D_m}\right)} - m\\
\tau_x\frac{dx}{dt} & = & m - x \\
\tau_T \frac{dV_T}{dt} & = & -(V_T-E_T)  +  D_T\sum_{\{\hat{t}_i\}}\delta(t-\hat{t}_i) 
\end{eqnarray}
where $I_s$ is the current injected in the soma, $I_d$ the current injected in the dendrites, $V_s$ is the somatic voltage, $V_d$ is the dendritic voltage, $m$ is the level of activation of a putative calcium current ($I_{\rm Ca}=g_1m$), $x$ is the level of activation of a putative calcium-activated potassium current  ($I_{\rm K(Ca)}=g_2x$), $V_T$ is the dynamic threshold for firing somatic spikes, $I_A$  is a spike-triggered current mediating adaptation, $I_{BAP}$ is the the current associated with the back-propagating action potential, $\epsilon_{sd}$ is the filter relating the current injected in the soma to the current arriving in the dendrite and $\epsilon_{ds}$ is the filter relating the current injected in the dendrite to the current arriving in the soma. The spikes are emitted if $V_s(t) > V_T(t)$ which results in $\hat{t}_{(last)} = t$ while $V_s \rightarrow E_r$ and $t \rightarrow t + \tau_R$.  The parameters are listed in Table \ref{Params}.
\begin{table} 
\begin{center}
\begin{tabular}{l|c|c|c}
\textbf{Variable} & &\textbf{Value} & \textbf{Units}  \\
\hline
Somatic leak conductance& $g_s$ & 22& nS  \\ 
Somatic capacitance& $C_s$ &379 & pF  \\
Somatic reversal potential&$E_s$ &-73 & mV \\
Threshold baseline &$E_T$ & -53 & mV \\
Spike-triggered jump in threshold&$D_T$ & 2.0 & mV  \\
Time-constant of dynamic threshold&$\tau_T$ & 27 &ms \\
Maximum `Ca' current &$g_1$ &567 & pA \\
Maximum effect of `Ca' current in soma&$\alpha$ & 337 & n.u.\\
Dendritic leak conductance &$g_d$ & 22& nS\\
Dendritic capacitance & $C_d$ & 86& pF\\
Dendritic reversal potential& $E_d$ & -53& mV\\
Time-constant for variable $m$& $\tau_m$&6.7& ms\\
Time-constant for variable $x$& $\tau_x $& 49.9 & ms\\
Sensitivity of `Ca' Current& $D_m$& 5.5 & ms\\
Maximum `K(Ca)' Current&$g_2$& -207& pA\\
Half-activtion potential of `Ca' current &$ E_m$& -0.6& mV\\
\end{tabular}
\caption{List of parameters and their fitted value for the two-compartment model.}\label{Params}
\end{center}
\end{table}

As a control, we also consider an entirely passive model of dendritic integration. In this model, the current injected in the dendrite is filtered passively to reach the soma. The generalized passive model has and instantaneous firing rate:
\begin{equation}
\lambda(t) = \lambda_0 \exp\left(\kappa_s *I_s + \kappa_{ds}*I_d + \sum_{\{\hat{t}_i\}}\eta_{A}(t-\hat{t}_i)\right)
\end{equation}
where $\lambda_0$ is a constant related to the reversal potential, $\kappa_s$ somatic membrane filter, $\kappa_{ds}$ is the filter relating the current injected in the dendrite to the voltage change in the soma, and $\eta_A$ is the effective spike-triggered adaptation.  

\subsection{Experimental Protocol}
Parasagittal brain slices of the somato-sensory cortex (300-350 m thick ) were prepared from 28-35 day-old Wistar rats. Slices were cut in ice-cold extracellular solution (ACSF), incubated at 34$^o$C for 20 min and stored at room temperature. During experiments, slices were superfused with in ACSF at 34$^o$C. The ACSF contained (in mM) 125 NaCl, 25 NaHCO3, 25 Glucose, 3 KCl, 1.25 NaH2PO4, 2 CaCl2 , 1 MgCl2 , pH 7.4, and was continuously bubbled with 5 \% CO2 / 95 \% O2. The intracellular solution contained (in mM) 115 K+-gluconate, 20 KCl, 2 Mg-ATP, 2 Na2-ATP, 10 Na2-phosphocreatine, 0.3 GTP, 10 HEPES, 0.1, 0.01 Alexa 594 and biocytin (0.2\%), pH 7.2.

Recording electrodes were pulled from thick-walled (0.25 mm) borosilicate glass capillaries and used without further modification (pipette tip resistance 5-10 M$\Omega$ for soma and 20-30M$\Omega$ for dendrites). Whole-cell voltage recordings were performed at the soma of a layer V pyramidal cell . After opening of the cellular membrane a fluorescent dye, Alexa 594 could diffuse in the entire neuron allowing to perform patch clamp recordings on the apical dendrite 600-700 $\mu$m from the soma. Both recordings were obtained using Axoclamp Dagan BVC-700A amplifiers (Dagan Corporation). Data was acquired with an ITC-16 board (Instrutech) at 10 kHz driven by routines written in the Igor software (Wavemetrics).

The injection waveform consisted of 6 blocks of 12 seconds. Each block is made of three parts: 1) one second of low-variance colored noise injected only in the soma, 2) one second of low-variance colored noise injected only in the dendritic injection site, 3) ten seconds of high-variance colored noise whose injection site depends on the block: In the first block, the 10-second stimulus is injected only in the dendritic site, the second block delivers the 10-second stimulus in the soma only, and the four remaining blocks deliver simultaneous injections in the soma and the dendrites. The colored noise was simulated with MATLAB as an Ornstein-Uhlenbeck process with a correlation time of 3 ms. The six blocks make a 72 seconds stimulus that was injected repeatedly without redrawing the colored noise (frozen-noise). Twenty repetitions of the 72-second stimulus were carried out, separated by periods of 2-120 seconds. Out of the twenty repetitions, a set of seven successive repetitions were selected on the basis of high intrinsic reliability.

\subsection{Fitting Methods}\label{SecFitMeth}
\begin{figure*}[t!]
\centering
\includegraphics[width = .98\textwidth]{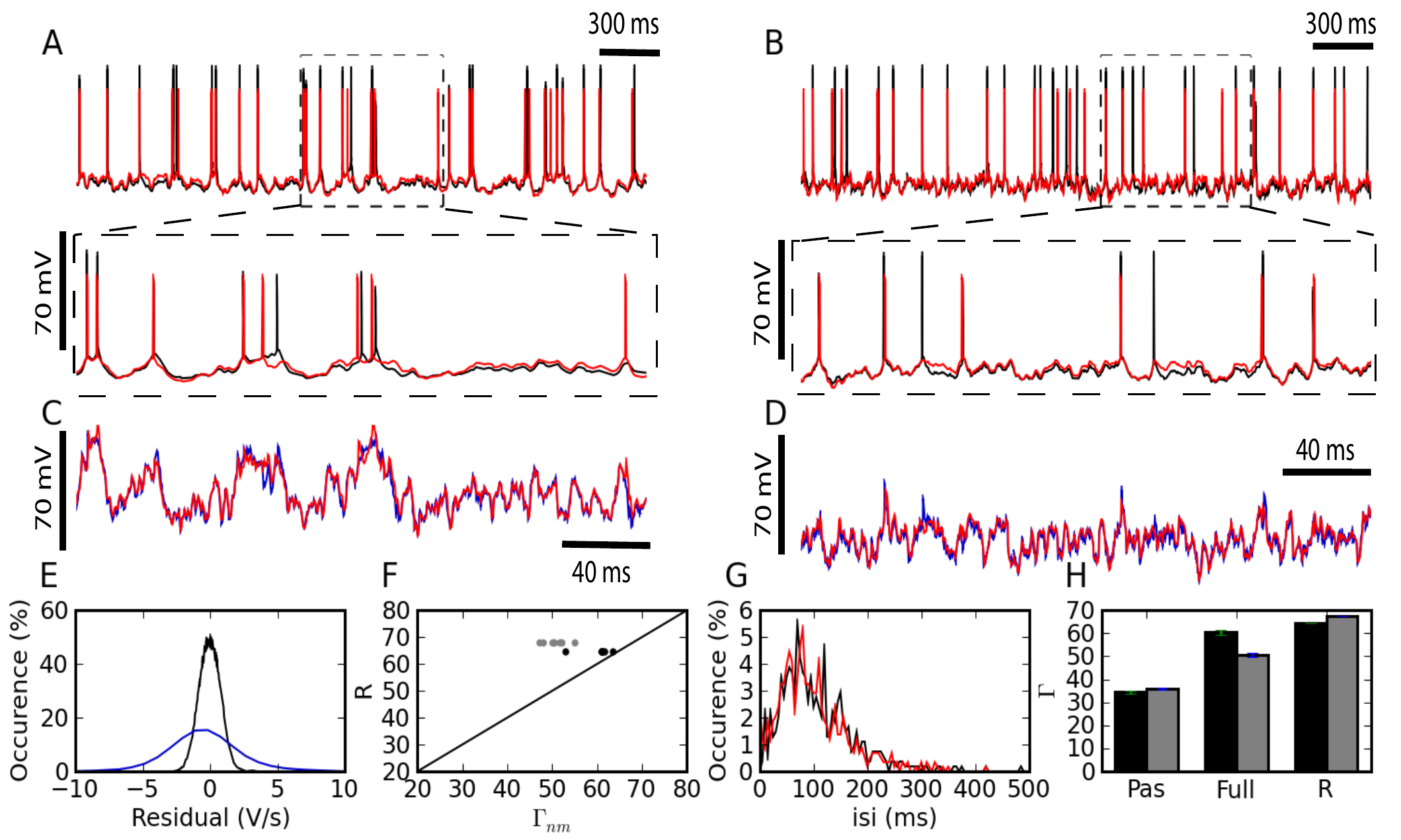} 
\caption{The two-compartment model fits qualitatively and quantitatively the electrophysiological recordings. {\sf\bf A, B } Overlay of the model (red) and experimental (black) somatic voltage trace.  The dashed box indicates an area stretched out for higher precision.   {\sf\bf C, D} The overlay of model (red) and experimental (blue) dendritic voltage is shown for the stretched sections in \textsf{\bf A} and \textsf{\bf B}.  Left ({\sf\bf A,C}) and right ({\sf\bf B,D}) columns show two different injection regimes contrasting by the amount of dendritic activity which is high for {\sf\bf A, C} and medium for {\sf\bf B, D}.  {\sf\bf E} Residuals from the linear regression are shown for the somatic (black) and dendritic (blue) compartment. {\sf\bf F} For each repetition the $\Gamma$ Coincidence factor is plotted against the intrinsic reliability of the cell.  Grey points show the performance of the model on the test set and black points show the performance of the model on the training set. {\sf\bf G} Comparison of the inter-spike interval histogram for the model (red) and the experiment (black).  {\sf\bf H} Comparison of the generalized passive (Pas), and the full two-compartment model (Full) with the intrinsic reliability (R) of the neuron in terms of the $\Gamma$ coincidence factor.  The averaged $\Gamma$ factor is shown for the training set (black) and test set (Gray)}\label{FigBig2Comp}
\end{figure*}

Each kernel ($\kappa_s$, $\kappa_{ds}$, $\eta_A$, $\epsilon_{ds}$,$\epsilon_{sd}$, $I_A$, $I_{BAP}$) is expressed as a linear combination of nonlinear basis (i.e. $\kappa_s(t) = \sum_i a_i f_i(t)$).  The rectangular function was chosen as the nonlinear basis. The parameters weighting the contributions of the different rectangular functions are then linear in the derivative of the membrane potential for the two-compartment model and generalized linear for the passive model.

For the two-compartment model, we use a combination of regression methods and exhaustive search to maximize the mean square-error of the voltage derivative. The regression methods are similar to those previously used for estimating parameters with intracellular recordings. These methods are described in more details in \cite{Jolivet2006a,Paninski2005a,Mensi2012a,Pozzorini2013a}. The fit of the somatic compartment essentially follows Jolivet \textit{et al.} (2006) \cite{Jolivet2006a} but using multi-linear regression to fit the linear parameters.  The fit of the dendritic compartment needs to iterate through the restricted set of nonlinear parameters ($\tau_m$, $D_m$, $E_m$, $\tau_x$). All fits are performed only on the part of the data restricted for training the model.

\begin{description}
\item[1] Fit of the dendritic compartment, knowing the injected currents and the somatic spiking history:
\begin{description}
\item[1a] Compute the first-order estimate of $dV_d/dt$;
\item[1b] Find the best estimates of the dendritic parameters linear in $dV_d/dt$ given a set of nonlinear parameters ($\tau_m$, $D_m$, $E_m$, $\tau_x$).  The best estimates are chosen through multi-linear regression to minimize the mean square error of $dV_d/dt$.
\item[1c] Compute iteratively step 1b on a grid of the nonlinear parameters ($\tau_m$, $D_m$, $E_m$, $\tau_x$) and find the nonlinear parameters that yield the minimum mean square error of $dV_d/dt$.
\end{description}
\item[2] Fit of the somatic compartment using the fitted dendritic compartment.
\begin{description}
\item[2a] Compute the first-order estimate of $dV_s/dt$. 
\item[2b]  Find the best estimates of the somatic parameters linear in $dV_s/dt$ given a set of nonlinear parameters ($D_T$, $\tau_T$, $E_T$).  The best estimates are chosen through linear regression to minimize the mean square error of $dV_d/dt$.
\item[2c] Compute iteratively step 2b on a grid of the nonlinear parameters and simulate the model with each set of nonlinear parameters in order to compute the coincidence rate $\Gamma$ (see Sect. \ref{sec-analysis}).
\item[2d] Take the parameters that yield the maximum $\Gamma$ coincidence factor.\end{description}
\end{description} 

For the generalized linear model, we use maximum likelihood methods \cite{Paninski2004b,Pillow2005a}. Expressing the kernels as a linear combination of rectangular bases we recover the generalized linear model. Here the link-function is exponential so that the likelihood is convex. We therefore performed a gradient ascent of the likelihood to arrive at the optimal parameters.

\subsection{Analysis Methods}
\label{sec-analysis}
When one focuses on spike timing, one may want to apply methods that compare spike trains in terms of a spike-train metric \cite{Victor1996a} or the coincidence rate \cite{Kistler1997a}. Both measures can be used to compare a recorded spike train with a model spike train. A model which achieve an optimal match in terms of spike-train metrics will automatically account for global features of the spike train such as the interspike interval distribution. 

Here we used the averaged coincidence rate $\Gamma$ \cite{Kistler1997a}. It can be seen as a similarity measure between pairs of spike trains, averaged on all possible pairs. To compute the pairwise coincidence rate, one first finds the number of spikes from the model that fall within an interval of $\Delta =$4~ms after or before a spike from the real neuron. This is called the number of coincident events $N_{nm}$ between neuron repetition $n$ and model repetition $m$. The coincidence rate is the ratio of the number of coincident events over the averaged number of events 0.5($N_n$+$N_m$), where $N_n$ is the number of spikes in the neuron spike train and $N_m$ is the number of spikes in the model spike train. This ratio is then scaled by the number of chance coincidences $N_{\rm Poisson} = 2\Delta N_m N_n/T$. This formula comes from the number of expected coincidences assuming a Poisson model at a fixed rate $N_m/T$ where $T$ is the time length of each individual spike trains. The scaled coincidence rate is 
\begin{equation}
\Gamma_{nm} = \frac{N_{nm} - N_{\rm Poisson}}{0.5(1-N_{\rm Poisson}/N_n)(N_n+N_m)}.
\end{equation}
The pairwise coincidence rate $\Gamma_{nm}$ is then averaged across all possible pairings of spike trains (trials) generated from the model with those from the neuron and gives the averaged coincidence rate $\Gamma$.  Averaging across all possible pairings of spike trains from the neuron with a distinct repetition of the same stimulus given to the same neuron gives the intrinsic reliability $R$.

\section{Results}
\label{sec-res}

Dual patch-clamp recordings were performed in L5 Pyramidal cells of Wistar rats (see Experimental Methods).  A simplified two-compartment model (see Model Description) was fitted on the first 36 seconds of stimulation for all repetitions. The rest of the data (36 sec) was reserved to evaluate the model's predictive power.  The predictive power of the two-compartment model with active dendrites was then compared to a model without activity in the dendrites (see Sect. \ref{ModDescrp}), the generalized linear passive model.  
\begin{figure}[h!]
\includegraphics[width = .5\textwidth]{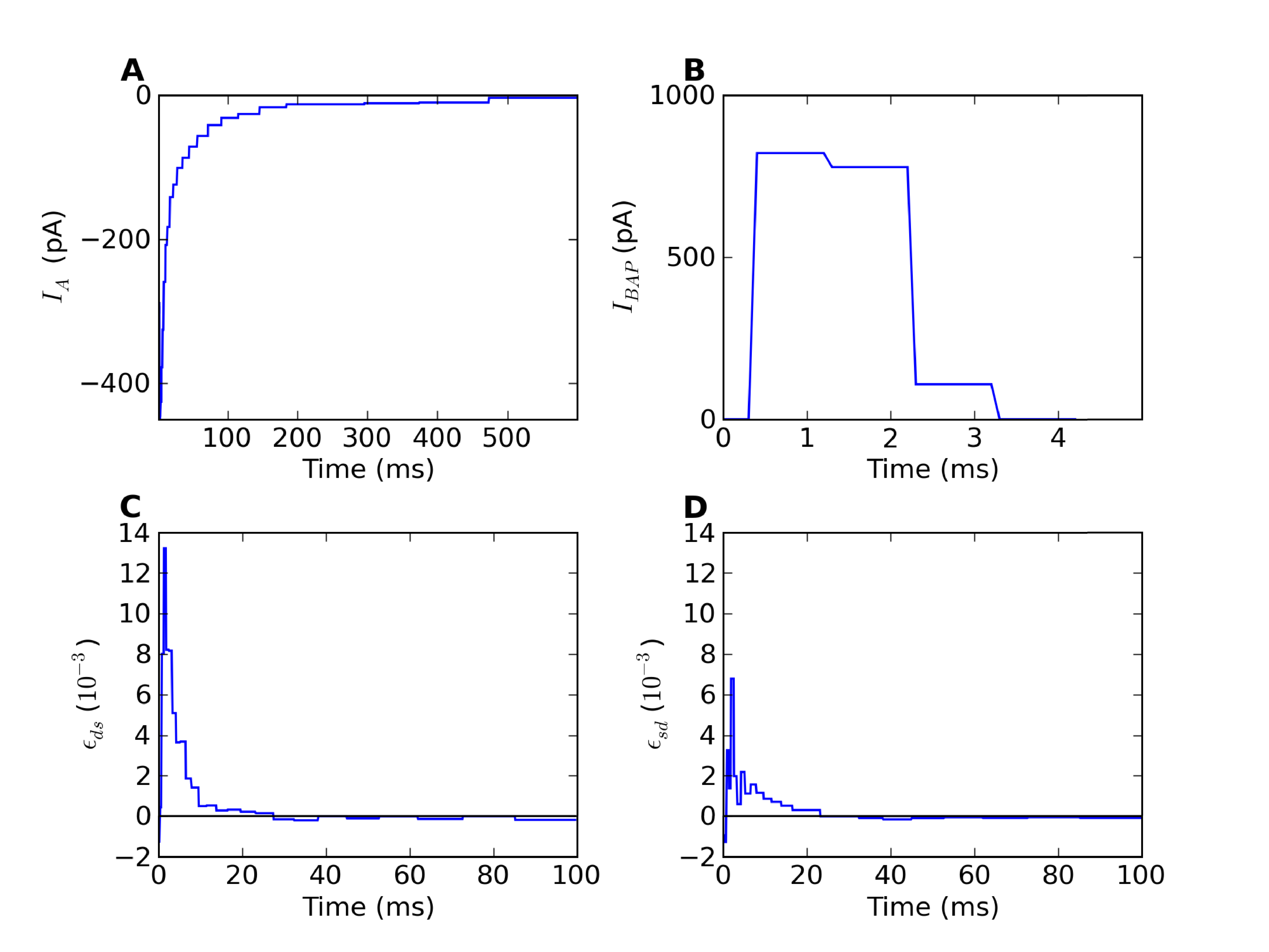} 
\caption{Fitted kernels of the two-compartment model.  {\bf A} The kernel $I_A(t)$ for spike-triggered adaptation is negative and increases monotonically between 6 and 600 ms. {\bf  B} The back-propagating current $I_{\rm BAP}(t)$reaching the dendrites is a short (2ms) and strong (900 pA) pulse. {\bf C}~The convolution kernel $\epsilon_{ds}(t)$ linking the current injected in the dendrite to the current reaching the soma.  {\bf  D} The convolution kernel $\epsilon_{sd}(t)$ linking the current injected in the soma to the current reaching the dendrite. }\label{Kernels}
\end{figure}

\begin{figure}
\centering
\includegraphics[width = .5\textwidth]{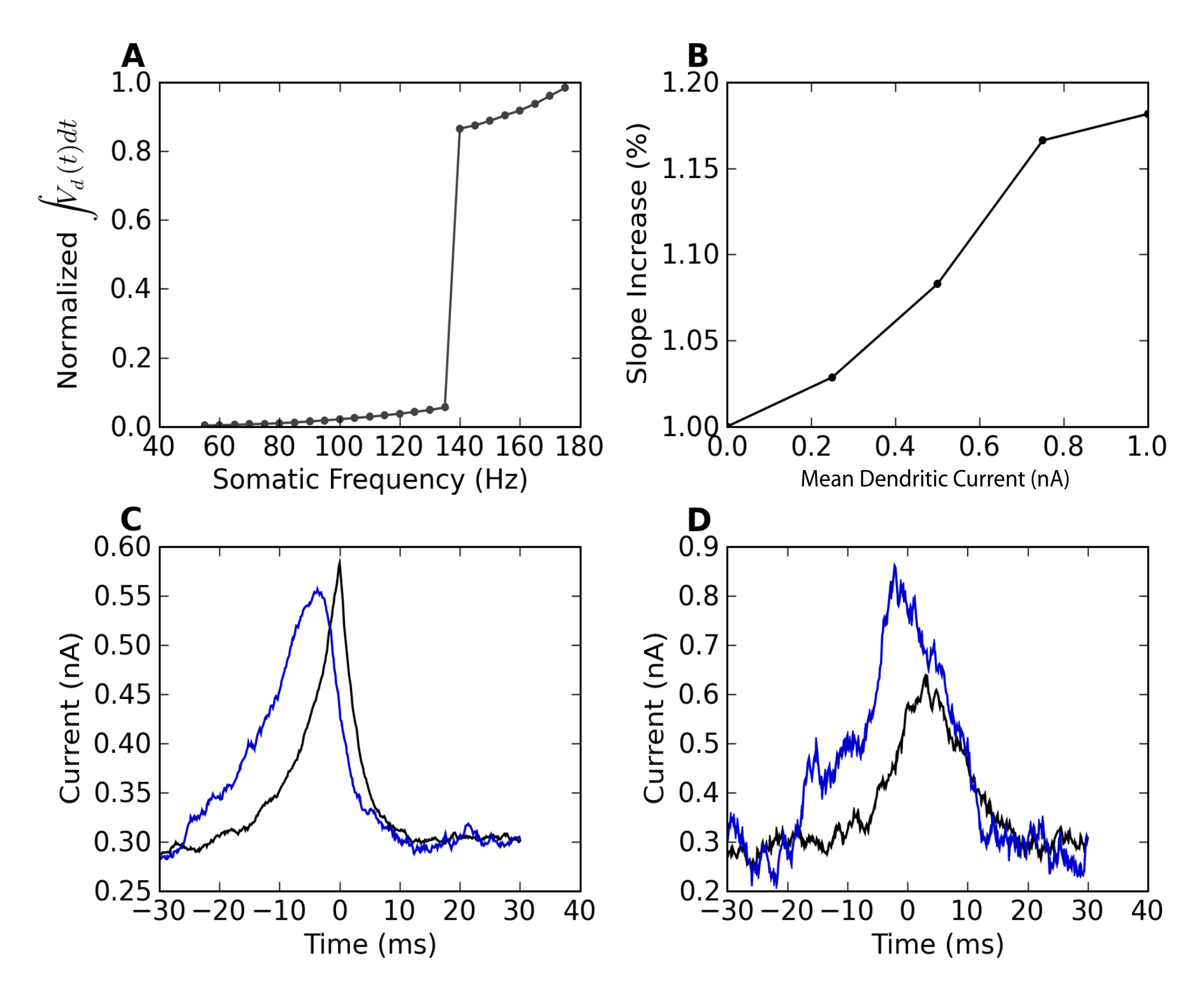}
\caption{The model reproduces the qualitative features of active dendrites reported in \cite{Larkum1999a} and \cite{Larkum2004a}. {\bf A} Dendritic non-linearity is triggered by somatic spiking above a critical frequency.  Somatic spike-trains of 5 spikes are forced in the soma of the mathematical model at different firing frequencies. The normalized integral of the dendritic voltage is shown as a function of the somatic spiking frequency.  {\bf B} Dendritic injection modulates the slope of the somatic spiking-frequency vs. current curve.  The slope of the frequency vs mean somatic current as measured between 5 and 50 Hz is plotted as a function of the mean dendritic current.  Both somatic and dendritic currents injected are Ornstein-Uhlenbeck processes with a correlation time of 3 ms and a standard deviation of 300 pA.  {\bf C} Spike-triggered average of the current injected in the soma (black) and in the dendrites (blue). {\bf D} Burst-triggered average of the current injected in the soma (black) and in the dendrites (blue).  The fact that the blue curve is higher than the black curve, and that this relation is inverted in C, indicate that the two-compartment model performs a type of top-down coincidence detection with bursts.
}\label{Pirouettes}
\end{figure}

Figure \ref{FigBig2Comp} summarizes the predictive power of the two-compartment model.  The somatic and dendritic voltage traces are well captured (Fig. \ref{FigBig2Comp} \textsf{\bf A}-\textsf{\bf D}).  The main cause for erroneous prediction of the somatic voltage trace is extra or missed spikes (Fig. \ref{FigBig2Comp} \textsf{\bf A} and \textsf{\bf B} lower panels).  The dendritic voltage trace of the model follows the recorded trace both in a low dendritic-input regime (Fig. \ref{FigBig2Comp} \textsf{\bf C}) and in a high dendritic-input regime with dendritic `spikes' (Fig. \ref{FigBig2Comp} \textsf{\bf D}).  The greater spread of voltage-prediction-error (Fig. \ref{FigBig2Comp}) is mainly explained by the larger range of voltages in the dendrites (somatic voltage prediction is strictly subthreshold whereas dendritic voltage prediction ranges from -70 mV to +40 mV).    The interspike interval distribution is well predicted by the model (Fig. \ref{FigBig2Comp} \textsf{\bf G}). \\

 The generalized passive model does not predict as many spike times ( Fig. \ref{FigBig2Comp} \textsf{\bf H}).  The intrinsic variability in the test set was 68\% and the two-compartment model predicted 50\%.  The prediction falls to 36 \% in the absence of a dendritic non-linearity (Fig. \ref{FigBig2Comp} \textsf{\bf H}).

The fitted kernels show that spike triggered adaptation is a monotonically decaying current that starts very strongly and decays slowly for at least 500 ms (Fig. \ref{Kernels} \textsf{\bf A}).  The back-propagating action potential is mediated by a strong pulse of current lasting 2-3 ms (Fig. \ref{Kernels} \textsf{\bf B}).  The coupling $\epsilon_{ds}$ from dendrite to soma has a maximal response after 2-3 ms and then decays so as to be slightly negative after 35 ms (Fig. \ref{Kernels} \textsf{\bf C}).  The coupling  $\epsilon_{sd}$ from soma to dendrite follows qualitatively $\epsilon_{ds}$ with smaller amplitudes and slightly larger delays for the maximum and minimum peaks (Fig. \ref{Kernels} \textsf{\bf D}), consistent with the larger membrane time-constant in the soma than in the dendrites.

The two-compartment model can reproduce qualitative features associated with the dendritic non-linearity in the apical tuft of L5 pyramidal neurons.  We study two of these features: the critical frequency \cite{Larkum1999a} and the gain modulation \cite{Larkum2004a}.  The first relates to the critical somatic firing frequency above which a non-linear response is seen in the soma, reflecting calcium channel activation in the dendrites.  To simulate the original experiment, we force 5 spikes in the soma at different frequencies and plot the integral of the dendritic voltage.  The critical frequency for initating a non-linear increase in summed dendritic voltage is 138 Hz (Fig. \ref{Pirouettes} \textsf{\bf A}). Perez-Garci \textit{et al.} (2006) \cite{Perez-Garci2006a} reported a critical frequency of 105 Hz while Larkum \textit{et al.} (1999) \cite{Larkum1999a} reported 85 Hz.  This appears to vary across different cells and pharmacological conditions.  

The model also appears to perform gain modulation as in \cite{Larkum2004a} (Fig. \ref{Pirouettes} \textsf{\bf B}).   The relation between somatic firing rate and mean somatic current depends on the dendritic excitability.  The onset (or shift) but also the gain (or slope) of the somatic frequency versus somatic current curve depend on the mean dendritic current. The gain modulation is attributed to a greater presence of bursts (Fig. \ref{Pirouettes} \textsf{\bf B}) caused by dendritic calcium-current activation at higher dendritic input. The link between burst and dendritic activity is reflected in the burst- and spike-triggered average injected current (Fig. \ref{Pirouettes} \textsf{\bf C}-\textsf{\bf D}) similar to Ref. \cite{Larkum2004a}.  The burst-triggered current is greater for the dendritic injection, whereas the spike-triggered current is larger for somatic injection. The greater correlation, relative to somatic current, of the dendritic current with the observation of bursts indicate that the two-compartment model performs a type of top-down coincidence detection with bursts.

\section{Conclusion}
Using a two-compartment model interconnected with temporal filters, we were able to predict a substantial fraction of spike times. The predicted spike trains achieved an averaged coincidence rate of 50\%. The scaled coincidence rate obtained by dividing by the intrinsic reliability \cite{Jolivet2008a,Naud2012b} was 72\%, which is comparable to the state-of-the performance for purely somatic current injection which reaches up to 76\%\cite{Naud2009a}.  Comparing with a passive model for dendritic current integration, we found that the predictive power decreased to a scaled coincidence rate of 53\%. Therefore we conclude that regenerating activity in the dendritic tuft is required to properly account for the dynamics of layer 5 pyramidal cells under in-vivo-like conditions.

\begin{acknowledgments}
The authors would like to thank Matthew Larkum for helpful suggestions.
\end{acknowledgments}


%

\end{document}